\documentclass[prl,twocolumn,showpacs, superscriptaddress
]{revtex4}
\newcommand{\be}{\begin{equation}}
\newcommand{\ee}{\end{equation}} 
\newcommand{\bea}{\begin{eqnarray}} 
\usepackage{graphicx}
\newcommand{\eea}{\end{eqnarray}}
\usepackage{epsfig}
\usepackage{bm}

\begin{document}
\title{Large Scale Intermittency in the Atmospheric Boundary Layer}
\author{M. Kholmyansky}
\affiliation{Faculty of Engineering, Tel Aviv University, Tel Aviv 69978, Israel}
\author{L. Moriconi}
\affiliation{Instituto de F\'\i sica, Universidade Federal do Rio de Janeiro, \\
C.P. 68528, 21945-970, Rio de Janeiro, RJ, Brazil}
\author{A. Tsinober}
\affiliation{Faculty of Engineering, Tel Aviv University, Tel Aviv 69978, Israel}
\affiliation{Marie Curie Chair in Fundamental and Conceptual Aspects of Turbulent
Flows, Institute for Mathematical Sciences and Department of Aeronautics,
Imperial College, London, 53 Princes Gate, Exhibition Road, South Kensington Campus,
London SW7 2PG, United Kingdom}

\begin{abstract} 
We find actual evidence, relying upon vorticity time series taken in a high Reynolds number atmospheric experiment, that to a very good approximation the surface boundary layer flow may be described, in a statistical sense and under certain regimes, as an advected ensemble of homogeneous 
turbulent systems, characterized by a lognormal distribution of fluctuating intensities. Our analysis suggests that usual direct numerical simulations of homogeneous and isotropic turbulence, performed at moderate Reynolds numbers, may play an important role in the study of turbulent boundary layer flows, if supplemented with appropriate statistical information concerned with the structure of large scale fluctuations.

\end{abstract}
\maketitle
A major difficulty in dealing with boundary layer flows at high Reynolds numbers is that they cannot be straightforwardly modeled within the theory of homogeneous isotropic turbulence, rendering unlikely, in principle, an application of the general results of the latter in the context of the former. 
In fact, essentially all the symmetry properties of the evolution equations break down close to the boundaries due to the intermittent production of a whole ``zoo" of flow phenomena, like low speed streaks and a number of vortex structures. The lack of homogeneity and isotropy is anagously observed far enough from the walls, where the transition to the outer laminar flow takes place, a region of strong entrainment and high intermittency factor, as discussed several decades ago by Klebanoff in his benchmark work (and subsequent papers) on turbulent boundary layers \cite{klebanoff}.

It has been long hypothesized, however, that in a typical turbulent boundary layer problem, as in the flow over a flat or rough surface, there is an intermediate range of normal distances from the boundary -- the logarithmic layer -- where the fundamental symmetries of the Navier-Stokes equation
are approximately restored at small scales, yielding a stage for tests of the statistical theory of turbulence. A recent investigation of related issues is provided by Sreenivasan et al \cite{sreenivasan}, from the analysis of the time series produced by hot-wire anemometry in a stable atmospheric boundary layer. It is worth noting that in atmospheric experiments it is usual to get samples where the flow is at best approximately statistically stationary, which makes the connection with the physics of homogeneous and isotropic turbulence not obvious at all. Actually, the experience shows that while the original velocity signal can be used to check the Kolmogorov's four-fifths law or the scaling exponents of structure functions, for instance, the lack of stationarity has to be carefully accounted for in the study of probability distributions of local fields. As a pragmatical solution of such a ``large scale intermittency" problem, the authors of Ref. \cite{sreenivasan} have retained from the rough anemometric data only the samples that would lead to statistically stationary regimes. It turned out, a posteriori, that their prescription worked consistently well.

The fact that several scaling features of homogeneous turbulence are observed without any further handling of the velocity signal seems to indicate that the underlying boundary layer flow could be modeled, in a first approximation which disregards shear effects, as an ensemble of homogeneous turbulent systems, collectively advected by the mean local velocity $U$. Each element of the ensemble would correspond to a flow with a definite value of the turbulent intensity $I \equiv u_{rms}/U$. Thus, if we are interested to model fluctuations of some observable defined at length scale $\ell$, $O(\ell,t)$, we may write
\be
O(\ell,t) = x(t) \tilde O(\ell,t) \ ,\ \label{oo}
\ee
where $\tilde O(\ell,t)$ denotes the observable fluctuation associated to an arbitrary homogeneous and isotropic turbulent flow, and $x(t)$ is an independent random function of time, which accounts for the fluctuations of the {\it{rms}} values of $O(\ell,t)$. In effect, $x(t)$ may be thought to play the role of a positive enveloping function which modulates the faster fluctuations of $O(\ell,t)$.
 
As an illustration of the kind of modeling we have in mind, take the case of longitudinal velocity differences $v_\ell(t) \equiv v_1 (\vec r+\ell \hat x,t)-v_1(\vec r,t)$. We have, from (\ref{oo}),
\be
v_\ell(t) = x(t) \tilde v_\ell(t) \ . \ \label{vv}
\ee
It is clear, therefore, that
\be
\langle v_\ell^q \rangle \equiv S_q(\ell) = \langle x^q \rangle \langle  \tilde v_\ell^q \rangle \propto \tilde S_q(\ell)  \ , \
\ee
that is, the structure functions $S_q(\ell)$ and $\tilde S_q(\ell)$ depend on the length scale $\ell$ exactly in the same way. On the other hand, it is not difficult to find that the probability distribution functions (pdfs) of $v_\ell$ and $\tilde v_\ell$ are, in general, completely different: we have
\be
\rho( v_\ell) = \int_0^\infty d x |x|^{-1} \tilde \rho( v_\ell / x) f(x)  \ , \ \label{pdf-bl}
\ee
where $f(x)$ is the pdf for the random variable $x$ introduced in (\ref{oo}), and $\rho( \cdot)$ and $\tilde \rho (\cdot)$ refer to the pdfs of velocity differences in boundary layer and homogeneous isotropic flows, respectively.
\begin{figure}[tbph]
\includegraphics[width=8.5cm, height=4.9cm]{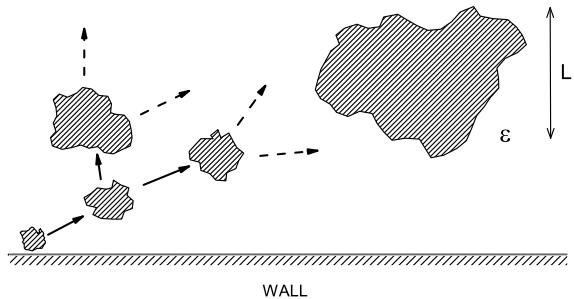}
\caption{Coherent structures generated close to the wall are transported to the bulk of the flow. These configurations are subject to random advection and instabilities causing them to disrupt, grow, and burst as the evolution proceeds. Approximately homogeneous and isotropic turbulence regions of size of order $L$ are produced with random energy transfer rates $\epsilon$.}
\label{fig1}
\end{figure}

Our central aim in this note is to discuss the statistics of the random enveloping function $x(t)$, through the analysis of the vorticity time series obtained from an atmospheric experiment carried out at very high Reynolds number \cite{kholm}. The measurements were performed with a 20 hot-wire probe, which incorporated specific design features appropriate for the particularities of the field experiment. The data was collected at a sampling rate of $10$ KHz (which was high enough to resolve the dissipative Kolmogorov scale) in a tower 10 m high, placed over a grass-covered flat surface. The total time length of the velocity signal is $15$ minutes, corresponding, by the mean wind velocity, to $6.3$ Km. The anemometer set was calibrated at the measurement position, to avoid possible perturbations caused by its manipulation and the variation of environmental factors. The Taylor-based Reynolds number of the flow, observed in approximately neutral and stable conditions, is $R_\lambda \simeq 10^4$; for a more detailed account of the experimental and phenomenological parameters, see \cite{kholm}. The experiment outcome, then, consists of time series for the three velocity components and all the nine components of the velocity-gradient tensor as well.

We intend to check, on the basis of purely heuristic and phenomenological arguments, whether the random variable $x$, as considered in Eq. (\ref{oo}), is lognormally distributed. A turbulent blob of size $L$ and rate of energy dissipation $\epsilon$ is produced in the boundary layer along a complex cascade directed towards larger scales, as shown in Fig.1. As a working hypothesis, an analogy with the multiplicative cascade arguments of the K62 phenomenology \cite{kolmogorov, obukhov} can be drawn here, assigning lognormal fluctuations to $\epsilon$. Its random behavior would be the result of sucessive disruptions of coherent vortex structures generated at the boundary, followed by their straining and transport to upper positions in the flow. Since $x(t)$ may be essentially identified to fluctuations of the $rms$ velocity, the usual expression
\be
u_{rms} \sim ( \epsilon L)^\frac{1}{3} \ , \ \label{urms}
\ee
taken from the statistical theory of turbulence \cite{m-y}, establishes a connection between $x(t)$ and $\epsilon$. Therefore, we expect that $u_{rms}$, or, equivalently, $x(t)$, will be lognormally distributed, at least approximately.

Manifestations of lognormal statistics in the physics of boundary layer flows are not unusual. It is worth mentioning that such distributions are found in turbulent boundary layers for: (i) the time intervals between velocity ``bursts" \cite{rao}; (ii) the spanwise separation between streaks intermittently produced at the wall \cite{flack}, and (iii) the small scale fluctuations of the dissipation field \cite{frehlich}. An interesting question is if all of these instances of lognormality can be related to each other within the framework of some unifying description.

Assuming that the $x$-variable in Eq. (\ref{oo}) is lognormally distributed is just half of the whole story. We also have to model the fluctuations of the small-scale observable of specific interest, which we take to be the vorticity field, $\omega_i = \epsilon_{ijk} \partial_j v_k$. There are experimental indications that even at moderate Reynolds numbers the enstrophy pdf (scaled to have unity variance) has, for a large range of enstrophy values, a turbulent asymptotic profile \cite{bersha_etal}. Relying on isotropy, that observation suggests that a similar asymptotic behavior should hold for the vorticity  fluctuations at Reynolds numbers which are not necessarily very high. In particular, it is likely that the vorticity pdfs obtained from direct numerical simulations, as the ones performed by Vincent and Meneguzzi \cite{vincent-meneguzzi} have asymptotic shapes (we mean simulations with $R_\lambda \sim 150$).

We have defined, motivated by the above considerations, a smooth polynomial interpolation of the Vincent-Meneguzzi numerical vorticity pdf and used that interpolation to generate a stochastic process for $\tilde \omega$, an arbitrary component of vorticity, by means of a standard Monte-Carlo procedure \cite{binder}. More precisely, let $i=1,2...$ be an integer index and $x(i)$ and $\tilde \omega (i)$ be independent random variables distributed according to the lognormal and the Vincent-Meneguzzi vorticity pdf, respectively, both with fixed variances. A stochastic process which simulates the vorticity fluctuations measured in the boundary layer is given by the series defined by 
\be
\omega(i) = x(i) \tilde \omega(i) \ . \ \label{stoc-proc}
\ee
The lognormality of the $x$-variable is implemented with the help of the mapping $x = \exp( y )$, where $y$ is normally distributed with variance $\sigma_y^2$. Observe that 
\be
H(n) \equiv \frac{ \langle  \omega^{2n} \rangle}{\langle  \omega^2 \rangle^n} = \frac{ \langle x^{2n} \rangle}{\langle x^2 \rangle^n} \cdot \frac{ \langle \tilde \omega^{2n} \rangle}{\langle \tilde \omega^2 \rangle^n}
= e^{ 2n(n-1) \sigma_y^2 } \tilde H(n) \ , \
\ee
a result that gives a hint on why the hyperflatness factors $H(n)$ are reasonable higher in boundary layers than the ones typically found in homogeneous and isotropic turbulent flows. We have considered a stochastic process with $31 \times 10^6$ elements, for various values of $\sigma_y$. The choice $\sigma_y = 0.51$ leads to an excellent agreement with the empirical vorticity pdfs, as shown in Fig.2.

\begin{figure}[tbph]
\hspace{-0.1cm}
{\centerline{
\includegraphics[width=10.0cm, height=8.6cm]{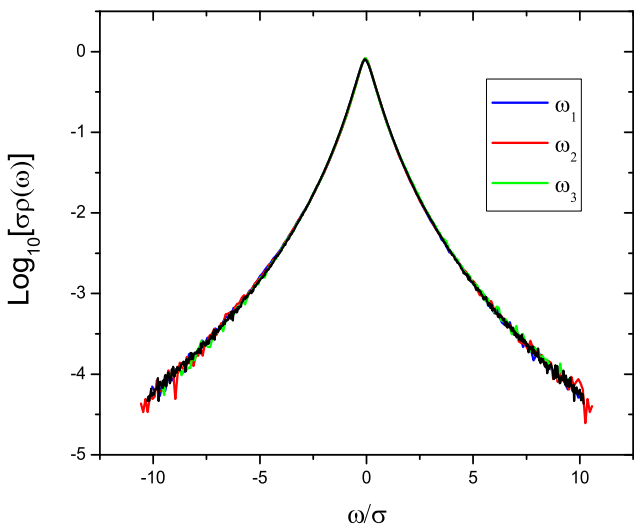}}}
\caption{Vorticity histograms. The black line is the result of the simulated stochastic process; the underlying colored lines are associated to the components of the vorticity vector, measured in the atmospheric boundary layer.}
\label{fig2}
\end{figure}

An important point concerns with the precision assigned to $\sigma_y$, and how it affects the vorticity distribution. We have found that alternative profiles for the pdf of $\tilde \omega$ and appropriate redefinitions of $\sigma_y$ lead to reasonable fittings as well. If, for example, $\tilde \omega$ is taken to be normally-distributed (which is obviously wrong) we get $\sigma_y = 0.7$. On the other hand, if we take a Student's t-distribution, which models reasonably well the vorticity pdf tails, as advanced in Ref. \cite{moriconi}, a very good agreement is found again, this time with $\sigma_y = 0.61$. The uncertainty in the value of $\sigma_y$ is partially due to properties of the lognormal distribution and partially related to the structure of Eq. (\ref{pdf-bl}). We will not discuss these mathematical aspects in detail, but it suffices to note that a random variable $x$, lognormally distributed, can be decomposed arbitrarily as the product of two other independent random variables, $x_1$ and $x_2$, both following lognormal distributions. This means that we may write
\be
x(i) = x_1(i) x_2(i)
\ee                                               
to get, from (\ref{stoc-proc}),
\be
\omega(i) =  x_1(i) [x_2(i) \tilde \omega(i)] = x_1(i) \bar \omega(i) \ , \ 
\ee                                                  
where $x_1(i)$  takes the place of  $x(i)$ and $\bar \omega(i) \equiv x_2(i) \tilde \omega(i)$ takes the place of $\tilde \omega(i)$ in Eq. (\ref{stoc-proc}). We find, thus, that the mere fitting of the vorticity pdf through the use of the stochastic process (\ref{stoc-proc}) has to be interpreted with care, even if the fitting is incredibly accurate. 

We conclude this note emphasizing that the problem of modeling fluctuations of vorticity -- or any other small-scale observable -- in the boundary layer is not so straightforward as it could seem at first sight. While we have demonstrated that the use of the lognormal distributions and statistical modeling based on DNS lead to excellent results, more elaborate statistical information is in order. A promising approach is to find instances where the same enveloping function $x(t)$, introduced in Eq. (\ref{oo}), would modulate intermittent fluctuations of alternative observables. We suggest filtered velocity fields as good candidates. Nevertheless the usual gaussian behavior of velocity, it is known that a frequency-filtering procedure applied to the velocity time series produces non-gaussian pdfs \cite{sreenivasan}, which, hopefully, could be modeled with the help of the lognormal random functions advanced here. Since this an issue of far more detailed analysis, a comprehensive discussion is deserved for future work. 

During the completion of this paper, we have become aware of an
article by F. Bottcher et al. \cite{bottcher}, which has
some overlap with the work presented here, mainly in the interpretation
of boundary layer flows in terms of homogeneous isotropic turbulent ensembles.
\acknowledgements
We thank Marie Farge for sending the numerical vorticity pdf files computed
from the Vincent-Meneguzzi DNS. L.M. thanks partial financial support by CNPq 
and FAPERJ.

\end{document}